\documentclass[%
reprint,
superscriptaddress,
nofootinbib,
nobibnotes,
amsmath,amssymb,
aps, prd
]{revtex4-2}

\usepackage{subcaption}
\usepackage{graphicx}
\usepackage{dcolumn}
\usepackage{bm}
\usepackage{hyperref}
\usepackage{amsmath} 
\usepackage{physics}
\usepackage{amssymb}
\usepackage{soul}
\usepackage{aas_macros}

\newcommand{\fDu}[1]{\stackon[-0.3ex]{$D^{#1}$}{\kern-1.0ex\scalebox{0.7}{$\circ$}}}

\newcommand{\zD}{{\raise1.0ex\hbox{${}^{\ \circ}$}}\!\!\!\!\!D}

\usepackage{orcidlink}

\graphicspath{{./}{figures/}{figures/res_1e6_pngs/}{figures/res_1e4_pngs/}{figures/res_1e3_pngs/}{figures/res_1e2_pngs/}}

\begin{document}

\preprint{APS/123-QED}

\title{Fundamental modes of rotating neutron stars with various degrees of differential rotation in dynamical spacetimes}

\author{Anson Ka Long \surname{Yip}~\orcidlink{0009-0008-8501-3535}}
\email{kalongyip@cuhk.edu.hk}
\affiliation{Department of Physics, The Chinese University of Hong Kong, Shatin, N.T., Hong Kong}

\author{Patrick Chi-Kit \surname{Cheong}~\orcidlink{0000-0003-1449-3363}}
\affiliation{Department of Physics, University of California, Berkeley, Berkeley, CA 94720, USA}
\affiliation{Center for Nonlinear Studies, Los Alamos National Laboratory, Los Alamos, NM 87545, USA}
\affiliation{Department of Physics \& Astronomy, University of New Hampshire, 9 Library Way, Durham NH 03824, USA}

\author{Tjonnie Guang Feng \surname{Li}~\orcidlink{0000-0003-4297-7365}}
\affiliation{Institute for Theoretical Physics, KU Leuven, Celestijnenlaan 200D, B-3001 Leuven, Belgium}
\affiliation{Department of Electrical Engineering (ESAT), KU Leuven, Kasteelpark Arenberg 10, B-3001 Leuven, Belgium }

\date{\today}

\begin{abstract}
Violent astrophysical events, including core-collapse supernovae and binary neutron star mergers, can result in rotating neutron stars with diverse degrees of differential rotation.
Oscillation modes of these neutron stars could be excited and emit strong gravitational waves. 
Detecting these modes may provide information about neutron stars, including their structures and dynamics.
Hence, dynamical simulations were employed to construct relations for quantifying the oscillation mode frequency in previous studies.
Specifically, linear relations for the frequencies of fundamental $l=0$ quasi-radial mode $f_{F}$ and fundamental $l=2$ quadrupolar mode $f_{^2f}$ were constructed by simulations with the Cowling approximation.
Nevertheless, these relations can overestimate $f_{F}$ and underestimate $f_{^2f}$ up to $\sim 30\%$.
Furthermore, it has yet to be fully studied how the degree of differential rotation affects $f_{F}$ and $f_{^2f}$.
Here, for the first time, we consider both various degrees of differential rotation $\Tilde{A}$ and dynamical spacetime to construct linear relations for quantifying $f_{F}$ and $f_{^2f}$.
Through 2D axisymmetric simulations, we first show that both $f_{F}$ and $f_{^2f}$ scale almost linearly with the stellar compactness $M/R$ for different values of $\Tilde{A}$.
We also observe the quasi-linear relations for both $f_{F}$ and $f_{^2f}$ with the kinetic-to-binding energy ratio $T/|W|$ for different $\Tilde{A}$ values.
Finally, we constructed linear fits that can quantify $f_{F}$ and $f_{^2f}$ by $T/|W|$.
Consequently, this work updated the relations for the fundamental modes of rotating neutron stars with differential rotations in dynamical spacetime.
\end{abstract}

\maketitle


\section{Introduction} \label{sec:intro}
Neutron stars with differential rotations are possible outcomes of astrophysical events, such as core-collapse supernovae and binary neutron star mergers (see e.g. \cite{2007PhR...442...38J,2016nure.book.....S,2017RPPh...80i6901B} for reviews).
The degree of differential rotation of these resulting neutron stars could vary in different events.
Core-collapse supernova simulations demonstrated that the ratio between central to 
equatorial angular velocity $\Omega_\mathrm{c}/\Omega_\mathrm{e}$ of the resulting proto-neutron stars is roughly 2 to 3 (see e.g. \cite{2004A&A...418..283V}).
On the other hand, for the hypermassive neutron stars formed in binary neutron star simulations, the angular velocity at the rotation axis could reach 1 order of magnitude higher than that at the surface (corresponds to $\Omega_\mathrm{c}/\Omega_\mathrm{e} \sim \mathcal{O}(10^{1})$) (see e.g. \cite{2008PhRvD..78h4033B}).

The neutron star remnants produced in these violent events are highly dynamical and their oscillation modes could be excited.
These excited oscillation modes could be detected in gravitational wave signals (see e.g. \cite{2004MNRAS.352.1089S,2006nwap.conf...25K} for detailed discussions).
Detecting these signals may provide us with information about neutron stars, such as the structure, dynamics, and equation of state (EOS) of the stars (see e.g. \cite{1998LRR.....1....8S} for a review of oscillations modes).
Oscillation modes of rotating neutron stars with and without differential rotations have been extensively studied by perturbative calculations or dynamical simulations \citep{1998MNRAS.299.1059A,1999ApJ...515..414Y,2001MNRAS.320..307K,2001MNRAS.325.1463F,2002ApJ...568L..41Y,2002PhRvD..65h4024F,2004PhRvD..70l4015B,2004MNRAS.352.1089S,2005MNRAS.356..217Y,2006MNRAS.368.1609D,2010PhRvD..81h4019K,2011PhRvD..83f4031G,2013PhRvD..88d4052D,2020PhRvL.125k1106K}.
In particular, \cite{2004MNRAS.352.1089S} studied the axisymmetric oscillation modes of uniformly rotating neutron stars and differentially rotating neutron stars in the Cowling approximation (keeping spacetime fixed while evolving matter equations).
The degree of differential rotation was fixed at $\hat{A} = A/r_\mathrm{e} = 1$ (or equivalently $\Tilde{A} = (A/r_\mathrm{e})^{-1}  = 1$ in this work) to mimick the rotational profiles of proto-neutron stars, where $r_\mathrm{e}$ is the equatorial radius of the star.
They found that the frequencies of both fundamental $l=0$ quasi-radial mode $f_{F}$ and fundamental $l=2$ quadrupolar mode $f_{^2f}$ of rotating neutron stars scale almost linearly with the kinetic-to-binding energy ratio $T/|W|$.
Then, by assuming a fixed relative differences between the actual frequencies and those obtained in their simulations in the Cowling approximation, \cite{2004MNRAS.352.1089S} also constructed  linear relations as functions of $T/|W|$ for predicting the actual frequencies of $f_{F}$ and $f_{^2f}$ of rotating neutron stars.
\cite{2006MNRAS.368.1609D} later re-examined this problem by including the effect of dynamical spacetime.
They demonstrated that simulations using the Cowling approximation can overestimate oscillation mode frequency up to a factor of 2.
They also found that the relative differences between the actual frequencies and those obtained in the Cowling approximation significantly change as $T/|W|$ increases.
As a consequence, applying linear relations derived from \cite{2004MNRAS.352.1089S} to predict actual frequencies can lead to an overestimate of $f_{F}$ and an underestimate of $f_{^2f}$ up to $\sim 30\%$.
Yet, \cite{2006MNRAS.368.1609D} also considered differentially rotating neutron stars with $\hat{A} = 1$ only.
For a better understanding of neutron star oscillations under various astrophysical scenarios, a study that takes into account various degrees of differential rotation and dynamical spacetime is required.

An ideal tool for conducting such a study would be dynamical simulations capable of examining a large number of stellar models in a reasonable amount of time and with minimal computational resources.
This kind of simulation can be performed using \texttt{Gmunu} \citep{2020CQGra..37n5015C,2021MNRAS.508.2279C,2022ApJS..261...22C}, a new general-relativistic magnetohydrodynamics code.
In \texttt{Gmunu}, simulations can be performed in multiple dimensions (1D, 2D, and 3D) as well as coordinate systems (cartesian, cylindrical, and spherical) using a block-based adaptive mesh refinement (AMR) module.
This flexibility allows users to select the most appropriate dimensionality and coordinate system for each problem, while also imposing symmetries to further reduce computational costs where possible.
For studies focused on axisymmetric ($m=0$) oscillation modes (e.g. $F$-mode and $^2f$-mode investigated by \cite{2004MNRAS.352.1089S,2006MNRAS.368.1609D}), 2D axisymmetric simulations performed in this study are particularly well-suited.
However, it is important to note that non-axisymmetric modes, such as the $m=1$ and $m=2$ modes, can arise and become significant in binary neutron star remnants and proto-neutron stars formed during core-collapse supernovae (see e.g. \cite{2015PhRvD..92l1502P}).
As a result, the assumption of 2D axisymmetry in this study hinders the capture of these non-axisymmetric modes in the simulations.
Investigating these modes would require 3D simulations without imposing axisymmetry.
Besides, using the multigrid method, \texttt{Gmunu} also solves the elliptic metric equations in the conformally flat condition (CFC) approximation efficiently and robustly.
All the features mentioned above make \texttt{Gmunu} an appropriate tool for studying problems related to a single neutron star (e.g. \cite{2021ApJ...915..108N,2022CmPhy...5..334L,2024MNRAS.534.3612Y,2023arXiv230515181Y,Yip:2023qkh}).

In this work, for the first time, we take into account both various degrees of differential rotation and dynamical spacetime to construct linear relations quantifying $f_{F}$ and $f_{^2f}$.
Specifically, we first construct initial neutron star models by the open-sourced code \texttt{XNS} \citep{2011A&A...528A.101B,2014MNRAS.439.3541P,2015MNRAS.447.2821P,2017MNRAS.470.2469P,2020A&A...640A..44S}.
A variety of differential rotation laws have been proposed in the literature to describe the differential rotation of protoneutron stars and binary neutron star merger remnants (see e.g \cite{2017PhRvD..96d3004H,2017PhRvD..96j3011U,2022MNRAS.510.2948I,2024MNRAS.532..945C}).
Similarly, realistic equations of state, including finite temperature effects, have been used to model these neutron star remnants (see e.g \cite{2017RPPh...80i6901B,2021PhRvD.103f3014C,2024PhRvD.110d3015C,2024PhRvD.110l4063M}).
However, the primary focus of this study is on the effects of degrees of differential rotation and dynamical spacetime rather than the effects coming from the details of the variety of rotation laws and equations of state discussed in the literature.
Therefore, following previous works \cite{2004MNRAS.352.1089S,2006MNRAS.368.1609D}, we adopt a simplified case of the $j$-constant differential rotation law \cite{1989MNRAS.237..355K,1989MNRAS.239..153K} and a polytropic equation of state, without considering finite-temperature effects or realistic equations of state, to construct the initial neutron star models.
While this approach allows us to isolate the qualitative impact of differential rotation and dynamical spacetime, these simplifying assumptions can lead to quantitative deviations in the computed mode frequencies. 
Thus, more realistic equations of state and rotation laws are necessary for a more accurate determination of the mode frequencies in scenarios such as protoneutron stars and binary neutron star merger remnants.
These equilibrium models are then perturbed and evolved in dynamical spacetime using \texttt{Gmunu}.
The details of the initial neutron star models and evolutions are described in Section \ref{sec:num_method}.
After that, we examine how the frequencies of fundamental modes vary with the compactness $M/R$ and the kinetic-to-binding energy ratio $T/|W|$ of neutron stars in Sections \ref{sec:freq_compact_r_ratio} and \ref{sec:freq_ratio} respectively.
Finally, we construct linear relations as functions of $T/|W|$ for quantifying $f_{F}$ and $f_{^2f}$ of rotating neutron stars with differential rotations and provide the conclusions in Section \ref{sec:conclusions}.

Unless otherwise specified, we choose dimensionless units for the physical quantities by setting the speed of light, the gravitational constant, and the solar mass to one, $c=G=M_\odot=1$.

\section{Numerical methods}\label{sec:num_method}
\subsection{Initial neutron star models}
We compute the neutron star equilibrium models in axisymmetry by the open-sourced code \texttt{XNS} \citep{2011A&A...528A.101B,2014MNRAS.439.3541P,2015MNRAS.447.2821P,2017MNRAS.470.2469P,2020A&A...640A..44S}. 
Our simulations are conducted using these models as initial data.

Initial neutron star models are constructed using a polytropic equation of state,
    \begin{equation}
    P=K \rho^\gamma,
    \end{equation}
where $P$ is the pressure, $\rho$ is the rest-mass density and we choose the polytropic constant $K =100$ and polytropic index $\gamma=2$. 

We set the specific internal energy $\epsilon$ on the initial time-slice by
    \begin{equation}
    \epsilon=\frac{K}{\gamma-1} \rho^{\gamma-1}.
    \end{equation}

The $j$-constant law introduced by \cite{1989MNRAS.237..355K,1989MNRAS.239..153K} has been widely used to model differentially rotating neutron stars (see e.g. \cite{2000ApJ...528L..29B,2004ApJ...610..941M,2014ApJ...790...19K} )

\begin{equation}
j(\Omega)=A^2\left(\Omega_{\mathrm{c}}-\Omega\right),
\end{equation}

where $j$ is the relativistic specific angular momentum, $A$ is a parameter to control the degree of differential rotation, $\Omega$ is the angular velocity and $\Omega_\mathrm{c}$ is the central angular velocity.  
Following previous studies (e.g. \cite{2010PhRvD..81h4019K}), we adopt the parameter $\Tilde{A} = (A/r_\mathrm{e})^{-1}$ to quantify the degree of differential rotation, where $r_\mathrm{e}$ is the equatorial radius of the equilibrium model.
Greater $\Tilde{A}$ corresponds to a higher degree of differential rotation and the star undergoes uniform rotation when $\Tilde{A}=0$.
We constructed equilibrium models in 5 sequences with $\Tilde{A} \in \{0.0,1.0,2.0,3.0,4.0\}$ to investigate the effect of the degree of differential rotation.
As we do not intend to investigate neutron stars with different masses in this work, we adopt a fixed baryonic mass $M_0=1.506$ for all models. 
3 neutron stars with high masses have been observed recently: J0348+0432 at $M = 2.01 \pm 0.04$ \citep{2013Sci...340..448A}, PSR J0740+6620 at $M = 2.08 \pm 0.07 $ \citep{2021ApJ...915L..12F}, and PSR J0952-0607 at $M = 2.35 \pm 0.17 $ \citep{2022ApJ...934L..17R}.
In light of these observations, a neutron star should have a maximum mass of at least $M = 2.0$. 
Accordingly, the masses of our models are within the observational constraints on the maximum mass of a neutron star.
Besides, the adopted baryonic mass $M_0=1.506$ is identical to that for the models in the previous study by \cite{2006MNRAS.368.1609D}, enabling comparison with that study.
The detailed properties of the equilibrium models are summarized in Appendix~\ref{sec_equil_models}.

\subsection{Evolutions}
Using the new general relativistic magnetohydrodynamics code \texttt{Gmunu} \citep{2020CQGra..37n5015C,2021MNRAS.508.2279C,2022ApJS..261...22C}, we evolve the stellar models in dynamical spacetime.
The models evolved over a time of 20 ms with the polytropic equation of state $P=K \rho^\gamma$, under the same conditions as equilibrium models (i.e. $K=100$ and $\gamma = 2$).

2D ideal general-relativistic hydrodynamics simulations are conducted in axisymmetry with respect to the $z$-axis and equatorial symmetry using spherical coordinates $(r,\theta)$. 
The computational domain covers $0 \leq r \leq 60$, $0 \leq \theta \leq \pi/2$, with the base grid resolution $N_{r} \times N_{\theta} = 64 \times 16$ and allowing 4 AMR levels (effective resolution = $512 \times 128$).
The refinement criteria of AMR is equivalent to that in \cite{2021MNRAS.508.2279C,2022CmPhy...5..334L,2024MNRAS.534.3612Y,2023arXiv230515181Y}.
TVDLF approximate Riemann solver \citep{1996JCoPh.128...82T} , 3rd-order reconstruction method PPM \citep{1984JCoPh..54..174C} and 3rd-order accurate SSPRK3 time integrator \citep{1988JCoPh..77..439S} are employed in our simulations.
Outside the star, there is an artificial atmosphere with rest-mass density $\rho_\mathrm{atm} \sim 10^{-10} \rho_\mathrm{c}$. 

\subsection{Initial perturbations}\label{sec:perturbations}
Following \cite{2006MNRAS.368.1609D}, we excite the fundamental modes by applying the following initial fluid perturbations on the equilibrium models.

First, we use the $l=0$ perturbation on the $r$-component of the three-velocity field $v^r$ to excite the fundamental $l=0$ quasi-radial mode (i.e. ${F}$-mode),
    \begin{equation}
        \delta v^r=a \sin\left[\pi\frac{r}{r_{\mathrm{s}}(\theta)}\right],
    \end{equation}
where $r_{\mathrm{s}}(\theta)$ denotes the radial position of the stellar surface, and the perturbation amplitude $a$ (in the unit of $c$) is chosen to be -0.005.

Second, we use the $l=2$ perturbation on the $\theta$-component of the three-velocity field $v^{\theta}$ to excite the fundamental $l=2$ quadrupolar mode (i.e. ${^2f}$-mode),
    \begin{equation}
        \delta v^{\theta}=a \sin\left[\pi\frac{r}{r_{\mathrm{s}}(\theta)}\right] \sin\theta \cos\theta,
    \end{equation}
where $a$ is chosen to be 0.01.

Fundamental modes are extracted by performing Fourier transforms of $v^{r}$ and $v^{\theta}$ at $r=3$, $\theta=\pi/4$, ensuring the extraction position is within the star (see Appendix~\ref{sec:mode_extract} or \cite{2006MNRAS.368.1609D} for more details).

\section{Fundamental mode frequencies against stellar compactness and radius ratio}\label{sec:freq_compact_r_ratio}
\begin{figure*}
    \centering
    \includegraphics[width=\textwidth, angle=0]{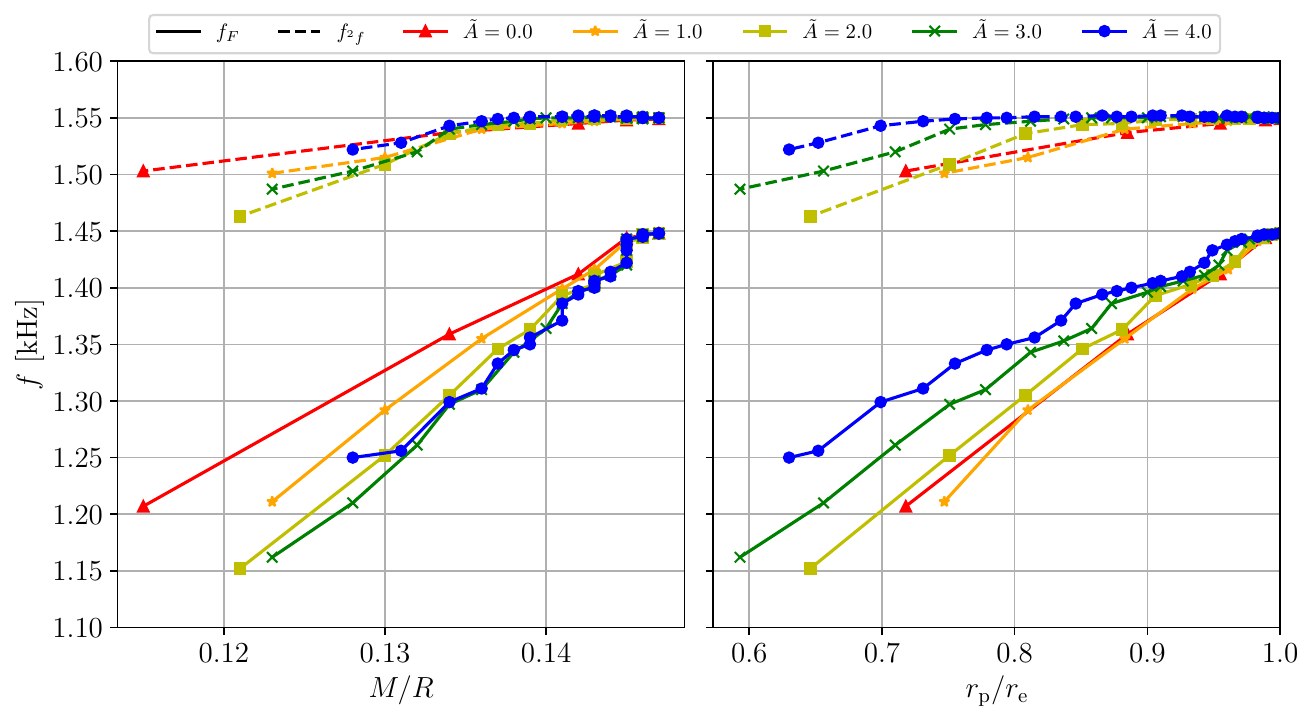}
    \caption{
             Plot of fundamental mode frequency $f$ against stellar compactness $M/R$ (left panel) and radius ratio $r_\mathrm{p}/r_\mathrm{e}$ (right panel), where $M$ is the stellar mass, $R$ is the circumferential radius, $r_\mathrm{p}$ is the polar radius and $r_\mathrm{e}$ is the equatorial radius.
             The data points are arranged into 5 sequences with $\Tilde{A} \in \{0.0,1.0,2.0,3.0,4.0\}$, where $\Tilde{A}$ is the degree of differential rotation.
             $\Tilde{A}=0.0$ refers to uniformly rotating models.
             Data points connected by solid lines denote the data for the frequency of fundamental $l=0$ quasi-radial mode $f_{F}$ while the points connected by dashed lines denote the data for the frequency of fundamental $l=2$ quadrupolar mode $f_{^2f}$.
             Both $f_{F}$ and $f_{^2f}$ increase approximately linearly with $M/R$ and $r_\mathrm{p}/r_\mathrm{e}$, with the slopes varying only slightly as $\Tilde{A}$ changes.
             Hence, this demonstrates a quasi-linear relation between fundamental mode frequency, stellar compactness and radius ratio for rotating neutron stars with different degrees of differential rotation.
            }
    \label{fig1}	
\end{figure*}

As the rotation rate of a neutron star increases, both the stellar compactness and radius ratio decrease, given that the baryonic mass remains constant (see e.g. \cite{2006MNRAS.368.1609D}).
\cite{1975ApJ...196..653H} have also shown that the frequencies of the fundamental modes are related to the stellar compactness of neutron stars.
To determine how how the fundamental mode frequency is related to both the stellar compactness and radius ratio in rotating neutron stars with different degrees of differential rotation, we plot the fundamental mode frequency $f$ against the stellar compactness $M/R$ (left panel) and radius ratio $r_\mathrm{p}/r_\mathrm{e}$ (right panel) in Fig.~\ref{fig1}, where $M$ is the stellar mass, $R$ is the circumferential radius, $r_\mathrm{p}$ is the polar radius and $r_\mathrm{e}$ is the equatorial radius.
The data points are arranged into 5 sequences with $\Tilde{A} \in \{0.0,1.0,2.0,3.0,4.0\}$, where $\Tilde{A}$ is the degree of differential rotation.
$\Tilde{A}=0.0$ refers to uniformly rotating cases.
Data points connected by solid lines represent the data for the frequency of fundamental $l=0$ quasi-radial mode $f_{F}$ while the points connected by dashed lines denote the data for the frequency of fundamental $l=2$ quadrupolar mode $f_{^2f}$.
The frequency $f$ of both fundamental modes increases with $M/R$ and $r_\mathrm{p}/r_\mathrm{e}$ almost linearly.
Slopes of $f(M/R)$ and $f(r_\mathrm{p}/r_\mathrm{e})$ for both fundamental modes changes only slightly with $\Tilde{A}$.
Hence, we observe a quasi-linear relation between fundamental mode frequency, stellar compactness and radius ratio for rotating neutron stars with different degrees of differential rotation.

\section{Fundamental mode frequencies against kinetic-to-binding energy ratio}\label{sec:freq_ratio}
\begin{figure*}
    \centering
    \includegraphics[width=\textwidth, angle=0]{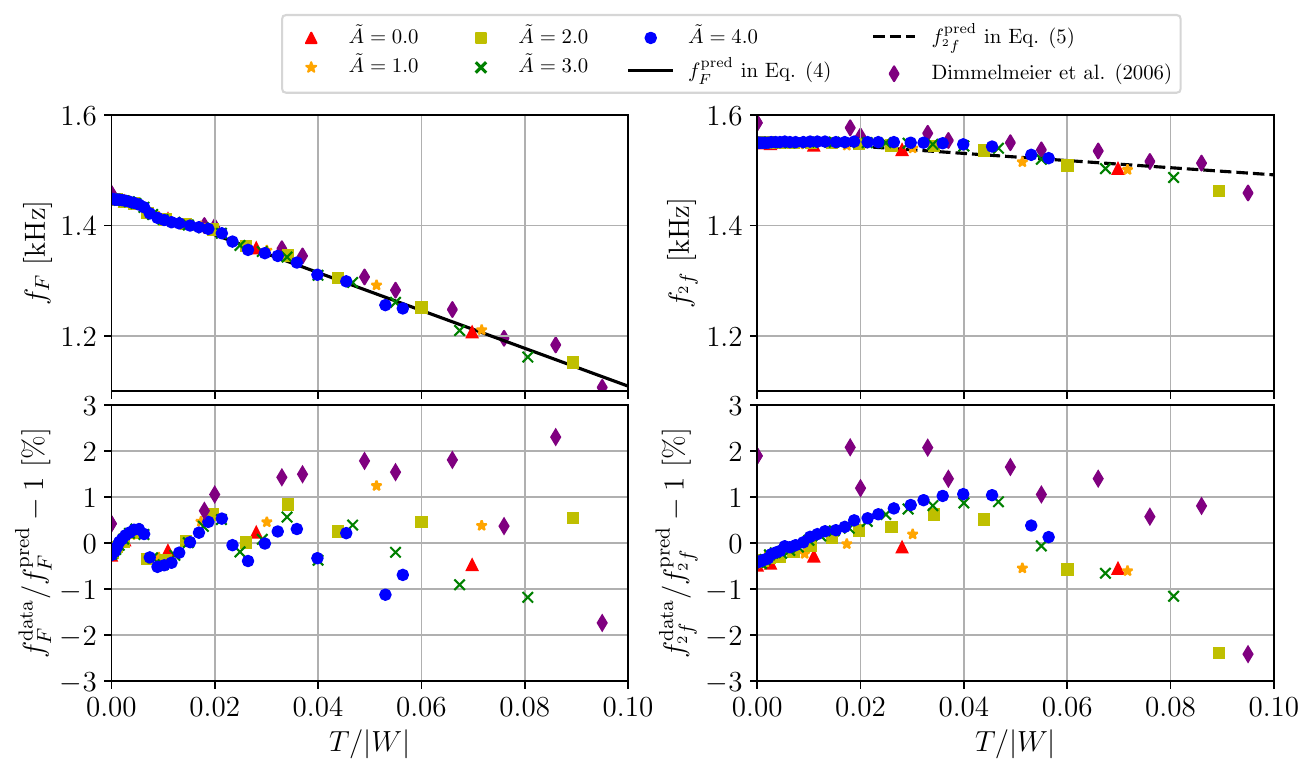}
    \caption{
             Plots of fundamental mode frequencies $f$ (top panels) and the frequency deviations between the simulation data $f^\mathrm{data}$ and the predictions by linear fits $f^\mathrm{pred}$ (bottom panels) against kinetic-to-binding energy ratio $T/|W|$.
             In particular, we plot fundamental $l=0$ quasi-radial mode frequency $f_{F}$ (top left panel), deviation of $f_{F}$ (bottom left panel), fundamental $l=2$ quadrupolar mode frequency $f_{^2f}$ (top right panel), and deviation of $f_{^2f}$ (bottom right panel) against $T/|W|$.
             The data points are arranged into 5 sequences with $\Tilde{A} \in \{0.0,1.0,2.0,3.0,4.0\}$, where $\Tilde{A}$ is the degree of differential rotation.
             $\Tilde{A}=0.0$ refers to uniformly rotating models.
             Both $f_{F}$ and $f_{^2f}$ for different $\Tilde{A}$ decrease almost linearly with $T/|W|$.
             Hence, we perform linear regressions using our simulation data to obtain 2 linear relations of $f^\mathrm{pred}_{F}(T/|W|)$ in Eq. (\ref{eqn6}) and $f^\mathrm{pred}_{^2f}(T/|W|)$ in Eq. (\ref{eqn7}) respectively.
             We find that only a slight deviation between our simulation data and our linear fits with $f^\mathrm{data}_{F}/f^\mathrm{pred}_{F} - 1 \lesssim 1 \%$ and $f^\mathrm{data}_{^2f}/f^\mathrm{pred}_{^2f} - 1 \lesssim 2 \%$.
             We also compare our linear fits with the data points of the previous study of \cite{2006MNRAS.368.1609D} (purple diamonds), in which the fundamental modes of uniformly rotating and differentially rotating ($\Tilde{A}=1.0$) neutron stars in dynamical spacetime were considered.
             We find that only minor deviations between the simulation data in \cite{2006MNRAS.368.1609D} and our linear fits with $f^\mathrm{data}/f^\mathrm{pred}- 1 \lesssim 2 \%$ for both fundamental modes.
             Therefore, we have constructed linear relations that can quantify $f_{F}$ and $f_{^2f}$ for rotating neutron stars with differential rotations in dynamical spacetime.}
    \label{fig2}	
\end{figure*}

As mentioned in Section \ref{sec:intro}, through simulations with the Cowling approximation, \cite{2004MNRAS.352.1089S} have demonstrated that the frequency of axisymmetric oscillation modes scales almost linearly with the kinetic-to-binding energy ratio for rotating neutron stars.
By assuming a fixed relative differences between the actual frequencies and those obtained in their simulations with the Cowling approximation, they have also constructed linear relations as functions of $T/|W|$ for predicting fundamental $l=0$ quasi-radial mode frequency $f_{F}$ and fundamental $l=2$ quadrupolar mode frequency $f_{^2f}$. 
Nonetheless, \cite{2006MNRAS.368.1609D} later showed that these linear relations can contribute to an overestimate of $f_{F}$ and underestimate of $f_{^2f}$ up to $\sim 30\%$.
In addition, both studies only regarded differentially rotating neutron stars with a single value of degree of differential rotation $\Tilde{A} = 1$.
Hence, we revisit the relation between the fundamental mode frequency $f$ and kinetic-to-binding energy ratio $T/|W|$ by considering both the effect of dynamical spacetime and various degrees of differential rotation.
We plot fundamental $l=0$ quasi-radial mode frequency $f_{F}$ (top left panel) and fundamental $l=2$ quadrupolar mode frequency $f_{^2f}$ (top right panel) against kinetic-to-binding energy ratio $T/|W|$ in Fig.~\ref{fig2}.
The data points are arranged into 5 sequences with $\Tilde{A} \in \{0.0,1.0,2.0,3.0,4.0\}$, where $\Tilde{A}$ is the degree of differential rotation.
$\Tilde{A}=0.0$ refers to uniformly rotating cases.
We observe that the frequencies $f$ of both fundamental modes for models with different $\Tilde{A}$ decrease roughly linearly with $T/|W|$.

Hence, using our simulation data, we perform linear regressions to obtain linear relations for quantifying fundamental $l=0$ quasi-radial mode frequency $f_{F}$ and fundamental $l=2$ quadrupolar mode frequency $f_{^2f}$ of rotating neutron stars with differential rotations.
For the quasi-radial $l=0$ fundamental mode frequency $f^\mathrm{pred}_{F}$,
        \begin{equation}\label{eqn6}
            f^\mathrm{pred}_{F}(\mathrm{kHz}) \approx 1.45-3.42 \frac{T}{|W|}.
        \end{equation}        
For the quadrupolar $l=2$ fundamental mode frequency $f^\mathrm{pred}_{^2f}$,
        \begin{equation}\label{eqn7}
            f^\mathrm{pred}_{^2f}(\mathrm{kHz}) \approx 1.56-0.65 \frac{T}{|W|}.
        \end{equation}        
We compute the deviations between our simulation data $f^\mathrm{data}$ and our linear fits $f^\mathrm{pred}$.
We then plot the deviations for $F$-mode $f^\mathrm{data}_{F}/f^\mathrm{pred}_{F} - 1$ (bottom left panel) and $^2f$-mode $f^\mathrm{data}_{^2f}/f^\mathrm{pred}_{^2f} -1 $ (bottom right panel) against $T/|W|$ in Fig.~\ref{fig2}.
We find that only a slight deviation between our simulation data and our linear fits with $f^\mathrm{data}_{F}/f^\mathrm{pred}_{F} - 1 \lesssim 1 \%$ and $f^\mathrm{data}_{^2f}/f^\mathrm{pred}_{^2f} - 1 \lesssim 2 \%$.
We also compare our linear fits with the data points of the previous study of \cite{2006MNRAS.368.1609D} (purple diamonds), in which the fundamental modes of uniformly rotating and differentially rotating ($\Tilde{A}=1.0$) neutron stars in dynamical spacetime were considered.
We find that only minor deviations between the simulation data in \cite{2006MNRAS.368.1609D} and our linear fits with $f^\mathrm{data}/f^\mathrm{pred}- 1 \lesssim 2 \%$ for both fundamental modes.
Thus, we have constructed linear relations that can quantify $f_{F}$ and $f_{^2f}$ of rotating neutron stars with differential rotations in dynamical spacetime.

We compare our linear relations with those constructed in the previous work with the Cowling approximation by \cite{2004MNRAS.352.1089S}.
For the predicted quasi-radial $l=0$ fundamental mode frequency $f_{F}$, they proposed
        \begin{equation}\label{eqn8}
            f_{F}(\mathrm{kHz}) \approx 1.44-2.59 \frac{T}{|W|}.
        \end{equation}         
For the predicted quadrupolar $l=2$ fundamental mode frequency $f_{^2f}$, they proposed
        \begin{equation}\label{eqn9}
            f_{^2f}(\mathrm{kHz}) \approx 1.58-3.69 \frac{T}{|W|}.
        \end{equation}     
In comparison to this work, our linear fit for quasi-radial $l=0$ fundamental mode frequency $f_{F}$ has a slightly smaller slope, while our linear fit for quadrupolar $l=2$ fundamental mode frequency $f_{^2f}$ has a much greater slope.
Accordingly, $f_{^2f}$ predicted by our linear fit decreases less significantly as the kinetic-to-binding energy ratio $T/|W|$ increases than the one predicted with the Cowling approximation.

These deviations are mainly due to the assumption of a fixed relative differences between the actual frequencies and those obtained in their simulation with the Cowling approximation.
In \cite{2004MNRAS.352.1089S}, they adopted an empirical relation between the actual frequency and the frequency in the Cowling approximation proposed by \cite{2002PhRvD..65h4024F} for constructing their linear fits,
\begin{equation}
f_{\Omega}=f_{\Omega}^{(\mathrm{C})}+\left[f_0-f_0^{(\mathrm{C})}\right]
\end{equation}
where the superscript (C) denotes the quantity in the Cowling approximation, $f_{\Omega}$ is the frequency of the fundamental mode for a rotating star with angular velocity $\Omega$ , and $f_0$ is the corresponding frequency in a non-rotating star.
This empirical relation was initially proposed based on uniformly rotating cases.
Hence, they assumed a fixed relative difference $f_0-f_0^{(\mathrm{C})}$ for all models.
As discussed in \cite{2006MNRAS.368.1609D}, the relative differences between the actual frequencies and those obtained in the Cowling approximation are significantly changed as the kinetic-to-binding energy ratio $T/|W|$ of the star increases.
Therefore, this assumption of a fixed relative difference would cause a deviation between our linear fits and those obtained in \cite{2006MNRAS.368.1609D}.

\section{Conclusions}\label{sec:conclusions}
In this work, for the first time, we considered various degrees of differential rotation $\Tilde{A}$ and dynamical spacetime to construct linear relations quantifying the frequencies of fundamental $l=0$ quasi-radial mode $f_{F}$ and fundamental $l=2$ quadrupolar mode $f_{^2f}$.
In particular, we first found that fundamental mode frequency $f$ scales nearly linearly with stellar compactness $M/R$.
Next, we observed quasi-linear relations between $f$ and the ratio of kinetic energy to binding energy $T/|W|$ for different values of $\Tilde{A}$.
Lastly, we constructed linear relations that can quantify $f_{F}$ and $f_{^2f}$ of rotating neutron stars with differential rotations based on linear regressions.
In consequence, these relations can quantify $f_{F}$ and $f_{^2f}$ for rotating neutron stars with differential rotations in dynamical spacetime.
 
Several natural extensions should be considered to model differentially rotating neutron stars in different scenarios more realistically.
First, for a more accurate determination of mode frequencies in neutron stars, particularly in scenarios such as core-collapse supernovae and neutron star mergers, it is important to use a more realistic equation of state that includes thermal effects (see e.g. \cite{2017RPPh...80i6901B,2021PhRvD.103f3014C}) as well as empirically motivated rotational laws (see e.g. \cite{2017PhRvD..96d3004H,2017PhRvD..96j3011U,2022MNRAS.510.2948I,2024MNRAS.532..945C,2024PhRvD.110d3015C,2024PhRvD.110l4063M}) as proposed in the literature for modeling these scenarios.
In addition, 3D simulations that do not assume axisymmetry should be performed to examine non-axisymmetric oscillation modes. 
These modes can arise and become significant in binary neutron star remnants and proto-neutron stars formed during core-collapse supernovae (see e.g. \cite{2015PhRvD..92l1502P}).
Furthermore, ultra-strong magnetic fields with strengths $B \sim \mathcal{O}(10^{17}) \mathrm{G}$ can be generated during the formation of proto-neutron stars \cite{2014MNRAS.439.3541P} and in binary neutron star mergers \cite{2006Sci...312..719P}. 
Such strong magnetic fields may also affect the frequencies of fundamental modes \cite{2022CmPhy...5..334L}. 
Therefore, the effects of magnetic fields should also be taken into account.

\begin{acknowledgments}
We acknowledge the support of the CUHK Central High-Performance Computing Cluster, on which the simulations in this work have been performed. 
This work was partially supported by grants from the Research Grants Council of Hong Kong (Project No. CUHK 14306419), the Croucher Innovation Award from the Croucher Foundation Hong Kong, and the Direct Grant for Research from the Research Committee of The Chinese University of Hong Kong. 
P.C.-K.C. acknowledges support from NSF Grant PHY-2020275 (Network for Neutrinos, Nuclear Astrophysics, and Symmetries (N3AS)).
\end{acknowledgments}

\appendix
\setcounter{table}{0}
\renewcommand{\thetable}{A\arabic{table}}

\setcounter{figure}{0}
\renewcommand{\thefigure}{A\arabic{figure}}

\section{Equilibrium models of neutron stars}\label{sec_equil_models}
This appendix summarizes the detailed properties of the equilibrium models of the 5 sequences in this study.
These data were used to construct the linear fits discussed in the main text.
We provide the data in the form of one table for each sequence (see Table~\ref{tablea1} - \ref{tablea5}).
Each sequence refers to each value of the degree of differential rotation $\Tilde{A} \in \{0.0,1.0,2.0,3.0,4.0\}$, where $\Tilde{A} = 0.0$ refers to uniform rotation.
In each table, we list the model, the central rest mass density $\rho_\mathrm{c}$, the gravitational mass $M$, the circumferential radius $R$, the equatorial radius $r_\mathrm{e}$, the ratio of polar radius to equatorial radius $r_\mathrm{p}/r_\mathrm{e}$, the central angular velocity $\Omega_\mathrm{c}$, the ratio of central angular velocity to equatorial angular velocity $\Omega_\mathrm{c}/\Omega_\mathrm{e}$, and the kinetic-to-binding energy ratio $T/|W|$.

\begin{table*}
	\centering
	\begin{tabular}{ccccccccc}
		 Model & $\rho_\mathrm{c}$ & $M$ & $R$ & $r_\mathrm{e}$ & $r_\mathrm{p}/r_\mathrm{e}$ & $\Omega_\mathrm{c}$ & $\Omega_\mathrm{c} / \Omega_\mathrm{e}$ & $T/|W|$\\
        & ($10^{-3}$) &   &   &   &   & ($10^{-2}$) &   &  \\
		\hline
  	 A0 & 1.280 & 1.400 & 9.547  & 8.086  & 1.000 & 0.000 & 1.000 & 0.000 \\
		 A1 & 1.264 & 1.401 & 9.642  & 8.180  & 0.989 & 0.500 & 1.000 & 0.003 \\
		 A2 & 1.214 & 1.402 & 9.881  & 8.414  & 0.955 & 1.000 & 1.000 & 0.011 \\
		 A3 & 1.117 & 1.406 & 10.453 & 8.977  & 0.885 & 1.500 & 1.000 & 0.028 \\
		 A4 & 0.909 & 1.414 & 12.300 & 10.805 & 0.718 & 2.000 & 1.000 & 0.070 \\
	\end{tabular}
	\caption{\label{tablea1} Stellar properties of the equilibrium models of sequence A constructed by the \texttt{XNS} code.
	A is a sequence of fixed rest mass $M_{0} = 1.506$ with the degree of differential rotation $\Tilde{A} = 0.0$ (i.e. uniform rotation).
	$\rho_\mathrm{c}$ is the central density, $M$ is the gravitational mass, $R$ is the circumferential radius, $r_\mathrm{e}$ is the equatorial radius, $r_\mathrm{p}/r_\mathrm{e}$ is the ratio of polar radius $r_\mathrm{p}$ to equatorial radius $r_\mathrm{e}$, $\Omega_\mathrm{c}$ is the central angular velocity, $\Omega_\mathrm{c} / \Omega_\mathrm{e}$ is the ratio of central angular velocity $\Omega_\mathrm{c}$ to equatorial angular velocity $\Omega_\mathrm{e}$, $T/|W|$ is the ratio of rotational kinetic energy $T$ to the absolute value of gravitational binding energy $|W|$.
    Numerical values of these stellar properties are rounded off to three decimal places.}
\end{table*}

\begin{table*}
	\centering
	\begin{tabular}{ccccccccc}
		 Model & $\rho_\mathrm{c}$ & $M$ & $R$ & $r_\mathrm{e}$ & $r_\mathrm{p}/r_\mathrm{e}$ & $\Omega_\mathrm{c}$ & $\Omega_\mathrm{c} / \Omega_\mathrm{e}$ & $T/|W|$\\
        & ($10^{-3}$) &   &   &   &   & ($10^{-2}$) &   &  \\
		\hline
		 B1 & 1.273 & 1.400 & 9.594  & 8.133  & 0.994 & 0.500 & 2.689 & 0.001 \\
		 B2 & 1.253 & 1.401 & 9.643  & 8.180  & 0.983 & 1.000 & 2.681 & 0.004 \\
		 B3 & 1.218 & 1.402 & 9.786  & 8.320  & 0.961 & 1.500 & 2.679 & 0.009 \\
		 B4 & 1.165 & 1.403 & 9.978  & 8.508  & 0.934 & 2.000 & 2.671 & 0.017 \\
   	 B5 & 1.086 & 1.406 & 10.313 & 8.836  & 0.883 & 2.500 & 2.645 & 0.030 \\
		 B6 & 0.963 & 1.410 & 10.888 & 9.398  & 0.810 & 3.000 & 2.618 & 0.051 \\
		 B7 & 0.855 & 1.414 & 11.508 & 10.008 & 0.747 & 3.250 & 2.585 & 0.072 \\
   
	\end{tabular}
	\caption{\label{tablea2} Stellar properties of the equilibrium models of sequence B constructed by the \texttt{XNS} code.
	B is a sequence of fixed rest mass $M_{0} = 1.506$ with the degree of differential rotation $\Tilde{A} = 1.0$.}
\end{table*}

\begin{table*}
	\centering
	\begin{tabular}{ccccccccc}
		 Model & $\rho_\mathrm{c}$ & $M$ & $R$ & $r_\mathrm{e}$ & $r_\mathrm{p}/r_\mathrm{e}$ & $\Omega_\mathrm{c}$ & $\Omega_\mathrm{c} / \Omega_\mathrm{e}$ & $T/|W|$\\
        & ($10^{-3}$) &   &   &   &   & ($10^{-2}$) &   &  \\
		\hline
		 C1  & 1.278 & 1.400 & 9.547  & 8.086  & 1.000 & 0.500 & 7.395 & 0.000 \\
		 C2  & 1.271 & 1.400 & 9.594  & 8.133  & 0.994 & 1.000 & 7.379 & 0.001 \\
		 C3  & 1.260 & 1.401 & 9.595  & 8.133  & 0.988 & 1.500 & 7.379 & 0.002 \\
		 C4  & 1.244 & 1.401 & 9.643  & 8.180  & 0.977 & 2.000 & 7.363 & 0.004 \\
   	 C5  & 1.224 & 1.401 & 9.691  & 8.227  & 0.966 & 2.500 & 7.348 & 0.007 \\
		 C6  & 1.197 & 1.412 & 9.787  & 8.320  & 0.949 & 3.000 & 7.319 & 0.010 \\
		 C7  & 1.165 & 1.413 & 9.836  & 8.367  & 0.933 & 3.500 & 7.305 & 0.014 \\
		 C8  & 1.126 & 1.404 & 9.979  & 8.508  & 0.906 & 4.000 & 7.301 & 0.020 \\
		 C9  & 1.079 & 1.405 & 10.123 & 8.648  & 0.881 & 4.500 & 7.300 & 0.026 \\
		 C10 & 1.019 & 1.407 & 10.269 & 8.789  & 0.851 & 5.000 & 7.205 & 0.034 \\
		 C11 & 0.951 & 1.409 & 10.508 & 9.023  & 0.808 & 5.500 & 7.252 & 0.044 \\
   	 C12 & 0.844 & 1.412 & 10.892 & 9.398  & 0.751 & 6.000 & 7.088 & 0.060 \\
		 C13 & 0.663 & 1.418 & 11.706 & 10.195 & 0.646 & 6.500 & 6.974 & 0.089 \\ 
	\end{tabular}
	\caption{\label{tablea3} Stellar properties of the equilibrium models of sequence C constructed by the \texttt{XNS} code.
	C is a sequence of fixed rest mass $M_{0} = 1.506$ with the degree of differential rotation $\Tilde{A} = 2.0$.}
\end{table*}

\begin{table*}
	\centering
	\begin{tabular}{ccccccccc}
		 Model & $\rho_\mathrm{c}$ & $M$ & $R$ & $r_\mathrm{e}$ & $r_\mathrm{p}/r_\mathrm{e}$ & $\Omega_\mathrm{c}$ & $\Omega_\mathrm{c} / \Omega_\mathrm{e}$ & $T/|W|$\\
        & ($10^{-3}$) &   &   &   &   & ($10^{-2}$) &   &  \\
		\hline
		 D1  & 1.279 & 1.400 & 9.547  & 8.086  & 1.000 & 0.500  & 15.187 & 0.000 \\
		 D2  & 1.276 & 1.400 & 9.547  & 8.086  & 1.000 & 1.000  & 15.187 & 0.000 \\
		 D3  & 1.272 & 1.400 & 9.594  & 8.133  & 0.994 & 1.500  & 15.165 & 0.001 \\
		 D4  & 1.265 & 1.400 & 9.594  & 8.133  & 0.988 & 2.000  & 15.164 & 0.002 \\
   	 D5  & 1.257 & 1.400 & 9.595  & 8.133  & 0.988 & 2.500  & 15.164 & 0.002 \\
		 D6  & 1.246 & 1.401 & 9.642  & 8.180  & 0.977 & 3.000  & 15.124 & 0.013 \\
		 D7  & 1.234 & 1.401 & 9.643  & 8.180  & 0.971 & 3.500  & 15.124 & 0.005 \\
		 D8  & 1.219 & 1.401 & 9.690  & 8.227  & 0.960 & 4.000  & 15.103 & 0.006 \\
		 D9  & 1.203 & 1.402 & 9.692  & 8.227  & 0.954 & 4.500  & 15.103 & 0.008 \\
		 D10 & 1.184 & 1.402 & 9.740  & 8.273  & 0.943 & 5.000  & 15.084 & 0.010 \\
		 D11 & 1.162 & 1.402 & 9.788  & 8.320  & 0.927 & 5.500  & 15.047 & 0.012 \\
   	 D12 & 1.138 & 1.403 & 9.836  & 8.367  & 0.910 & 6.000  & 15.030 & 0.015 \\
		 D13 & 1.111 & 1.403 & 9.884  & 8.414  & 0.900 & 6.500  & 14.995 & 0.018 \\
   	 D14 & 1.080 & 1.404 & 9.980  & 8.508  & 0.873 & 7.000  & 14.948 & 0.021 \\
		 D15 & 1.047 & 1.405 & 10.029 & 8.555  & 0.858 & 7.500  & 14.917 & 0.025 \\
		 D16 & 1.008 & 1.406 & 10.125 & 8.648  & 0.837 & 8.000  & 14.859 & 0.029 \\
   	 D17 & 0.966 & 1.407 & 10.222 & 8.742  & 0.812 & 8.500  & 14.822 & 0.034 \\
		 D18 & 0.916 & 1.408 & 10.365 & 8.883  & 0.778 & 9.000  & 14.749 & 0.040 \\
		 D19 & 0.859 & 1.409 & 10.510 & 9.023  & 0.751 & 9.500  & 14.669 & 0.047 \\
		 D20 & 0.791 & 1.411 & 10.702 & 9.211  & 0.710 & 10.000 & 14.580 & 0.055 \\
		 D21 & 0.687 & 1.414 & 11.037 & 9.539  & 0.656 & 10.500 & 14.425 & 0.067 \\
		 D22 & 0.560 & 1.416 & 11.515 & 10.008 & 0.593 & 10.750 & 14.324 & 0.081 \\

	\end{tabular}
	\caption{\label{tablea4} Stellar properties of the equilibrium models of sequence D constructed by the \texttt{XNS} code.
	D is a sequence of fixed rest mass $M_{0} = 1.506$ with the degree of differential rotation $\Tilde{A} = 3.0$.}
\end{table*}

\begin{table*}
	\centering
	\begin{tabular}{ccccccccc}
		 Model & $\rho_\mathrm{c}$ & $M$ & $R$ & $r_\mathrm{e}$ & $r_\mathrm{p}/r_\mathrm{e}$ & $\Omega_\mathrm{c}$ & $\Omega_\mathrm{c} / \Omega_\mathrm{e}$ & $T/|W|$\\
        & ($10^{-3}$) &   &   &   &   & ($10^{-2}$) &   &  \\
		\hline
		 E1  & 1.279 & 1.400 & 9.547  & 8.086  & 1.000 & 0.500  & 25.230 & 0.000 \\
		 E2  & 1.278 & 1.400 & 9.547  & 8.086  & 1.000 & 1.000  & 25.230 & 0.000 \\
		 E3  & 1.275 & 1.400 & 9.547  & 8.086  & 1.000 & 1.500  & 25.230 & 0.001 \\
		 E4  & 1.272 & 1.400 & 9.594  & 8.133  & 0.994 & 2.000  & 25.456 & 0.002 \\
   	 E5  & 1.268 & 1.400 & 9.594  & 8.133  & 0.988 & 2.500  & 25.456 & 0.002 \\
		 E6  & 1.262 & 1.400 & 9.594  & 8.133  & 0.988 & 3.000  & 25.456 & 0.013 \\
		 E7  & 1.256 & 1.400 & 9.594  & 8.133  & 0.983 & 3.500  & 25.456 & 0.005 \\
		 E8  & 1.248 & 1.401 & 9.595  & 8.133  & 0.983 & 4.000  & 25.456 & 0.006 \\
		 E9  & 1.239 & 1.401 & 9.642  & 8.180  & 0.971 & 4.500  & 25.111 & 0.008 \\
		 E10 & 1.229 & 1.401 & 9.643  & 8.180  & 0.966 & 5.000  & 25.111 & 0.010 \\
		 E11 & 1.219 & 1.401 & 9.643  & 8.180  & 0.960 & 5.500  & 25.111 & 0.012 \\
   	 E12 & 1.207 & 1.401 & 9.690  & 8.227  & 0.949 & 6.000  & 25.333 & 0.015 \\
		 E13 & 1.195 & 1.401 & 9.691  & 8.227  & 0.943 & 6.500  & 25.334 & 0.018 \\
   	 E14 & 1.179 & 1.402 & 9.739  & 8.273  & 0.932 & 7.000  & 25.001 & 0.021 \\
		 E15 & 1.164 & 1.402 & 9.740  & 8.273  & 0.926 & 7.500  & 25.001 & 0.025 \\
		 E16 & 1.147 & 1.402 & 9.787  & 8.320  & 0.910 & 8.000  & 25.220 & 0.029 \\
   	 E17 & 1.131 & 1.403 & 9.788  & 8.320  & 0.904 & 8.500  & 25.221 & 0.034 \\
		 E18 & 1.108 & 1.403 & 9.836  & 8.367  & 0.888 & 9.000  & 24.900 & 0.040 \\
		 E19 & 1.089 & 1.403 & 9.884  & 8.414  & 0.877 & 9.500  & 25.117 & 0.047 \\
		 E20 & 1.068 & 1.404 & 9.885  & 8.414  & 0.866 & 10.000 & 25.116 & 0.055 \\
		 E21 & 1.041 & 1.404 & 9.980  & 8.508  & 0.846 & 10.500 & 25.020 & 0.067 \\
   	 E22 & 1.018 & 1.405 & 9.981  & 8.508  & 0.835 & 11.000 & 25.020 & 0.055 \\
		 E23 & 0.986 & 1.405 & 10.076 & 8.602  & 0.815 & 11.500 & 24.931 & 0.067 \\
   	 E24 & 0.952 & 1.406 & 10.125 & 8.648  & 0.794 & 12.000 & 24.641 & 0.055 \\
		 E25 & 0.923 & 1.406 & 10.174 & 8.695  & 0.779 & 12.500 & 24.848 & 0.067 \\
   	 E26 & 0.882 & 1.407 & 10.269 & 8.789  & 0.755 & 13.000 & 24.772 & 0.055 \\
		 E27 & 0.836 & 1.408 & 10.365 & 8.883  & 0.731 & 13.500 & 24.699 & 0.067 \\
   	 E28 & 0.770 & 1.409 & 10.509 & 9.023  & 0.699 & 14.000 & 24.373 & 0.055 \\
		 E29 & 0.668 & 1.410 & 10.795 & 9.305  & 0.652 & 14.500 & 24.216 & 0.067 \\
		 E30 & 0.610 & 1.411 & 10.984 & 9.492  & 0.630 & 14.750 & 24.538 & 0.067 \\

	\end{tabular}
	\caption{\label{tablea5} Stellar properties of the equilibrium models of sequence E constructed by the \texttt{XNS} code.
	E is a sequence of fixed rest mass $M_{0} = 1.506$ with the degree of differential rotation $\Tilde{A} = 4.0$.}
\end{table*}

\section{\label{sec:mode_extract}Mode extraction and identification}
\begin{figure*}
    \centering
    \includegraphics[width=\textwidth, angle=0]{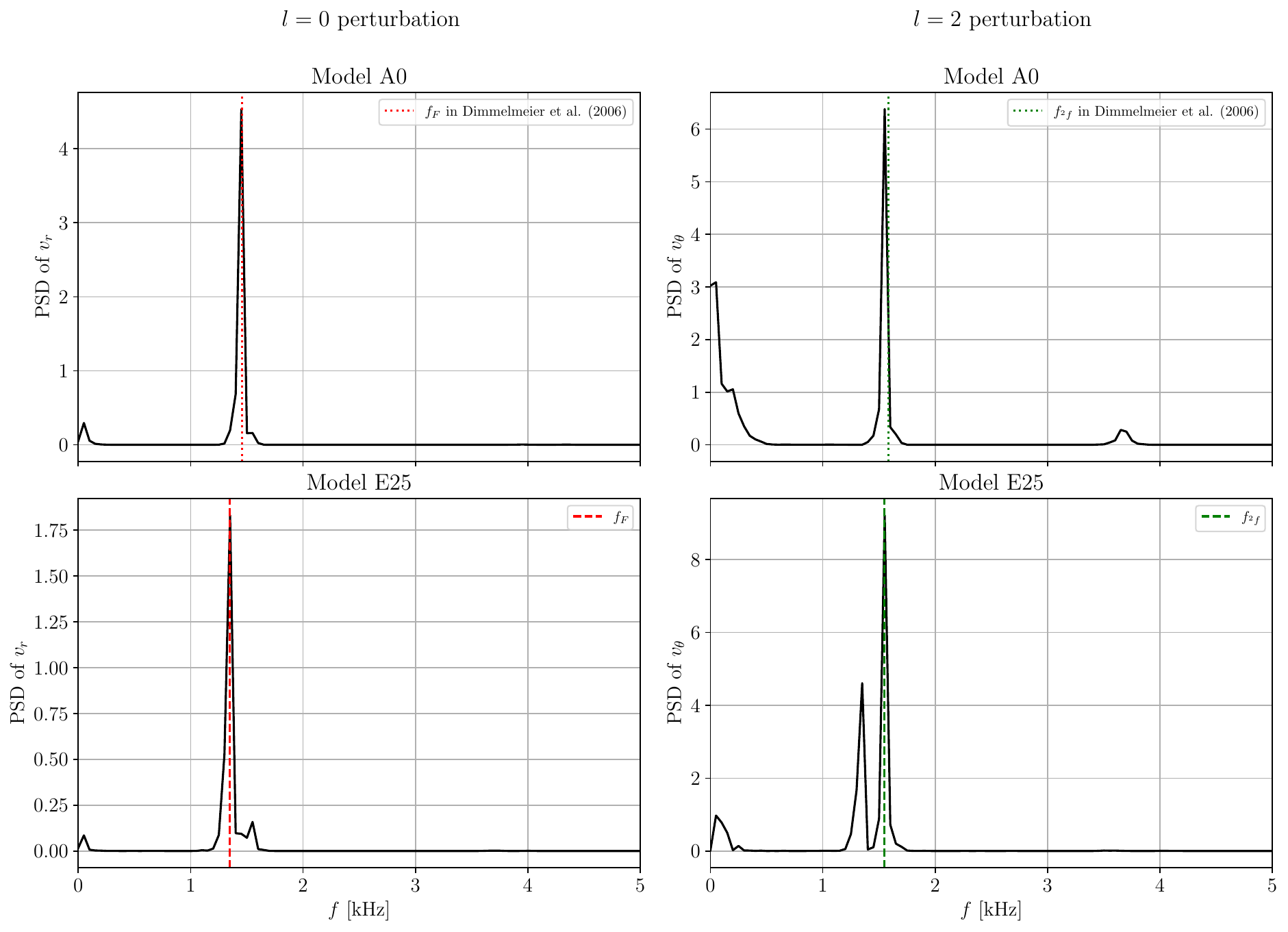}
    \caption{The power spectral density (PSD) of the $r$-component $v_r$ with $l=0$ perturbation imposed (left panels) and the $\theta$-component $v_\theta$ with $l=2$ perturbation imposed (right panels) of the three-velocity field in arbitrary units for the non-rotating model A0 (top panels) and model E25 (bottom panels).
    We first identify the peaks in the PSDs that correspond to the fundamental $l=0$ quasi-radial mode frequency $f_F$ and the fundamental $l=2$ quadrupolar mode frequency $f_{^2f}$ for our model A0 (top panels) by comparing them with the well-tested mode frequencies reported by Dimmelmeier et al. (dotted lines).
    After identifying the peaks of $f_F$ and $f_{^2f}$ in the PSDs, we track how $f_F$ and $f_{^2f}$ change in each sequence as the kinetic-to-binding energy ratio $T/|W|$ increases, as illustrated in the example of model E25 (bottom panels).}
    \label{figa1}	
\end{figure*}

To illustrate how we identify and extract the fundamental modes, we plot the power spectral density (PSD) of the $r$-component $v_r$ with an $l=0$ perturbation imposed (left panels) and the $\theta$-component $v_\theta$ with an $l=2$ perturbation imposed (right panels) of the three-velocity field in arbitrary units for the non-rotating model A0 (top panels) and model E25 (bottom panels) in Fig.\ref{figa1}.
Since our non-rotating model A0 corresponds to ‘BU0’ (or ‘AU0’) in the previous study by Dimmelmeier et al.\cite{2006MNRAS.368.1609D}, we can identify the peaks in the PSDs of $v_r$ and $v_\theta$ that correspond to the fundamental $l=0$ quasi-radial mode frequency $f_F$ and the fundamental $l=2$ quadrupolar mode frequency $f_{^2f}$ for our model A0 by comparing them with the well-tested mode frequencies reported by Dimmelmeier et al. (dotted lines), as shown in the top panels of Fig.\ref{figa1}.
After identifying the peaks of $f_F$ and $f_{^2f}$ in the PSDs, we track how $f_F$ and $f_{^2f}$ change in each sequence as the kinetic-to-binding energy ratio $T/|W|$ increases.
As shown in the bottom panels of Fig.\ref{figa1}, we take model E25, which is one of the models in the sequence with the highest degree of differential rotation ($\Tilde{A}=4.0$), as an illustrative example to demonstrate how we extract $f_F$ and $f_{^2f}$ (dashed lines) from the PSDs.

\bibliography{references}{}

\end{document}